\documentclass[%
  twoside,
  floatfix,
  reprint,
  amsmath,amssymb,
  aps,
  pra,
  nofootinbib,
  showpacs,
  superscriptaddress,
  a4paper
]{revtex4-1}

\usepackage{graphicx}% 
\usepackage[usenames,dvipsnames]{xcolor}
\usepackage{siunitx}
\usepackage{subfigure}
\usepackage{enumitem}
\usepackage{subfigure}

\usepackage{tabularx}
\usepackage{booktabs}
\usepackage{multirow}

\usepackage{titlesec}

\usepackage{array}

\usepackage[utf8]{inputenc}
\usepackage[T1]{fontenc}

\usepackage{lipsum}

\graphicspath{{Figures/}{}}

\usepackage[
centering, includefoot,
text={7.1in,10.2in},
total={6.3in,8.75in},
top=0.8in, left=0.62in,
]{geometry}

\usepackage[
  bookmarks=false,
  colorlinks,
  linkcolor=blue,
  urlcolor=blue,
  citecolor=blue,
  plainpages=false,
  pdfpagelabels,
  final,
  breaklinks=true
]{hyperref}
\hypersetup{
pdftitle={Laser-assisted photoionization of argon atoms: streaking, sideband and pulse train studying cases}, 
pdfauthor={R. Della Picca, M. F. Ciappina, M Lewenstein and D. Arb\'o}
}
\usepackage{natbib}
\makeatletter \def\NAT@def@citea{\def\@citea{\NAT@separator\,}} \makeatother

\newcommand{\orcid}[1]{%
  \href{%
    https://orcid.org/#1%
  }{%
   \,\protect\includegraphics[width=8pt]{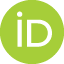}%
  }%
}

\usepackage{physics}
  \newcommand{\vbr}{\vb{r}}
  \newcommand{\vbp}{\vb{p}}
  \newcommand{\vbk}{\vb{k}}
  \newcommand{\vba}{\vb{A}}
  \newcommand{\vbf}{\vb{F}}
  \newcommand{\vbd}{\vb{d}}
  
  \newcommand{\vbv}{\vb{v}}
 \newcommand{\rme}{\mathrm{e}}
\newcommand{\rmd}{\mathrm{d}}

\newcommand{\Eref}[1]{Equation (\ref{#1})}
\newcommand{\eref}[1]{Eq.~(\ref{#1})}

\newcommand{\infi}{\ensuremath{\mathrm{if}}}

\DeclareSIUnit{\au}{{a.u.}}

\newcommand{\nhphantom}[1]{\sbox0{#1}\hspace{-\the\wd0}}
\begin{document}
\title{Laser-assisted photoionization of argon atoms:\\
streaking, sideband, and pulse train studying cases}

\author{R. Della Picca\,\orcid{0000-0001-7909-4529}}
\email[]{renata@cab.cnea.gov.ar}
\affiliation{Centro At\'omico Bariloche (CNEA),  CONICET and Instituto Balseiro (UNCuyo), 8400 Bariloche, Argentina}

\author{M. F. Ciappina\,\orcid{0000-0002-1123-6460}}
\affiliation{ICFO -- Institut de Ciencies Fotoniques, The Barcelona Institute of Science and Technology, 08860 Castelldefels (Barcelona)}

\author{Maciej Lewenstein\,\orcid{0000-0002-0210-7800}\,}
 \affiliation{ICFO -- Institut de Ciencies Fotoniques, The Barcelona Institute of Science and Technology, 08860 Castelldefels (Barcelona)}
 \affiliation{ICREA, Passeig de Llu\'is Companys, 23, 08010 Barcelona, Spain}

\author{D. G. Arb\'o\,\orcid{0000-0002-4375-4940}}
\affiliation{Institute for Astronomy and Space Physics - IAFE (UBA-Conicet), Buenos Aires, Argentina}
\affiliation{Universidad de Buenos Aires - Facultad de Ciencias Exactas y Naturales y Ciclo B\'asico Com\'un , Buenos Aires, Argentina}

\date{\today}

\begin{abstract}
We present a theoretical study of atomic laser-assisted photoionization emission (LAPE).
We consider an atom driven by a linearly polarized XUV laser in two different scenarios: i) a single attosecond pulse (in both the streaking and sideband regimes) and ii) an attosecond pulse train. The process takes place \textit{assisted} by a linearly polarized infrared (IR) laser field.
In all these cases the energy and angle-resolved photoelectron spectrum (PES) is determined by a leading contribution, related to the \textit{intracycle} factor [Gramajo \textit{et al.}, J. Phys. B {\bf 51}, 055603 (2018)], complemented by other ones, derived from the periodicity and symmetry properties of the dipole transition matrix with respect to the IR field. Each of these terms imprint particular features in the PES that can be straightforwardly understood in terms of generalized energy conservation laws.
We investigate in detail these PES structures, in particular, for the case of 
argon initially in the 3s quantum state. Our theoretical scheme, based on the strong-field approximation (SFA), can be applied, however, to other atomic species and field configurations as well.

\end{abstract}

\pacs{32.80.Wr, 32.80.Fb, 03.65.Sq}
\maketitle

%\preprint{APS/123-QED} %\tableofcontents

%-------------------------------------------------------------
\section{Introduction}
%-------------------------------------------------------------

Laser-assisted photoionization emission (LAPE) processes take place when extreme ultraviolet (XUV) radiation and infrared (IR) intense laser fields overlap in space and time. 
Two different scenarios arise depending on the XUV pulse duration: the streaking regime, if the XUV pulse is shorter than one IR optical cycle, and the sideband regime, if the XUV pulse is longer. In the first case, an electron wavepacket is put into the continuum by the XUV pulse in the presence of the IR laser field. Provided that the fields of these two pulses are controlled with sub-fs temporal resolution, the photoelectron spectra for different delays between the pulses, referred to as spectrograms, contain information about both the amplitude and phase of both the XUV and IR fields. Applying a reconstruction algorithm, these parameters can be straightforwardly retrieved~\cite{Mairesse2005,Goulielmakis2008,Goulielmakis2004,Gagnon2009}.

%Since the first theoretical prediction of ``sidebands" peaks~\cite{Veniard95}, an ample amount of experiments and theoretical studies have been performed in this area, see e.g.~\cite{Itatani02, Drescher05, Maquet2007, Radcliffe2012, Meyer2012, Mazza2014,Dusterer2019} and references therein.
On the other hand, in the second scenario, the simultaneous absorption of one high-frequency photon, together with the exchange of several additional photons from the IR laser field, leads to  equally spaced ``sideband'' peaks in the energy-resolved photoelectron spectra (PES), located on each side of the XUV photoionization energy value \cite{Drescher2005,Meyer2010}. 
Since the first theoretical prediction of ``sidebands'' peaks~\cite{Veniard95}, an ample amount of experiments and theoretical studies have been performed in this area, see e.g.~\cite{Itatani02, Drescher05, Maquet2007, Radcliffe2012, Meyer2012, Mazza2014,Dusterer2019} and references therein.
From the theoretical point of view, the formation of these peaks can be equivalently explained as the constructive interference between electron wavepackets emitted at different optical cycles % periods 
of the IR laser field~\cite{Kazansky10b,Gramajo18}. 

Experimentally speaking, the production of a train of attosecond pulses is easier than an isolated attosecond burst generation~\cite{Paul2001}. An attosecond pulse train synchronized with an IR laser pulse may assist a delay-dependent photoionization probability as well as probe the dissociative ionization of small molecules (e.g.~H$_2$)~\cite{schafer2007,Cocke2010,Toshima2010,Vrakking2011}. Furthermore, copies of the nuclear wavepacket can be produced by an attosecond pulse train, during molecular ionization. These replicas, however, are prone to be incoherently summed up, because of the entanglement between the laser-ionized electron and its parent molecular ion~\cite{Thumm2010}. Recently, a combination of a circularly polarized laser field and a train of XUV pulses was employed to extract the carrier envelope phase of the latter, analyzing the interference patterns that show up in the photoelectron momentum distributions~\cite{He2016}.

Within the context of laser-assisted potential scattering, it has been shown that the differential cross-section for the collision process, accompanied with the positive (absorption) or negative (stimulated emission) exchange of photons from the dressing field, can be factorized as a field-free term and a function that accounts for the laser field, via the classical excursion vector of a free electron and the peak amplitude of the laser electric field~\cite{KrollWatson}.  In this approach, dubbed `soft-photon' approximation, it is assumed that the photon energy of the laser field that ‘dresses’ the atomic continuum states is substantially smaller than the kinetic energy of the photoelectron. The soft-photon approximation can be adapted to laser-assisted photoionization, under the condition that the electron is freed by the XUV field, meanwhile the IR only acts 'dressing' the electron continuum and does not play any role in the laser-ionization process~\cite{Maquet2007}. 

There exists two general nonperturbative approaches that are nowadays widely used in strong-field atomic and molecular physics. The first one is based on the stationary treatment of the time-dependent Schr\"odinger equation (TDSE). Here, the so-called generalized Floquet formalisms allows the reduction of the periodical or quasiperiodical TDSE into a set of time-independent coupled equations, also known as the Floquet matrix eigenvalue problem. Floquet methods have been applied to an ample range of atomic and molecular multiphoton and tunneling processes in the last three decades. The initial limitations of Floquet-like methods, however, have been already lifted, allowing stationary treatment of laser pulse excitation problems (see e.g.~\cite{Chu2004} and references therein).
The second scheme is to solve numerically the TDSE, discretising both the time and spatial coordinates. The advantage of the time-dependent approaches is that they can be applied directly to many problems, ranging from multiphoton excitation to tunneling ionization, and for fields of arbitrary shape and duration. The main drawback, however, is the high computational cost, particularly for long wavelength sources~\cite{Qprop}.

In previous works~\cite{Gramajo18, Gramajo16,Gramajo17} we have employed a semiclassical model (SCM), based on the strong field approximation (SFA), to identify the electron trajectories and describe the energy and angle-resolved photoelectron spectrum (PES)
as the product of inter- and intracycle interferences factors. The former accounts for the sidebands' formation and the latter appears as a modulation of them. Additionally, we have also shown that it is possible to write the PES as a function of the time dependent photoionization transition matrix for an XUV pulse in the presence of one IR cycle~\cite{proceedingIcpeac19}. 
These interferences were derived using the saddle point approximation in the temporal integration of the transition matrix. 

In this work we describe the PES in a more general way without resorting to the saddle point approximation. To this end, we explore the photoionization of argon atoms for different configurations of the XUV laser field, \textit{assisted} by an IR field. Specifically,
we consider ionization by a single attosecond pulse, in both the streaking and sideband regimes and, additionally, the case of an XUV pulse train. 
High resolution experiments, under the mentioned field arrangements, %conditions, 
would be desirable in order to confirm the PES structures identified in
%predictions of the 
the present study. 
Unlike other models, such as those based on the Floquet theory or the soft photon approximation, both originally proposed for infinitely long pulses, the present approach is theoretically correct for any duration of both the IR and XUV pulses. One additional advantage, as every approximation with roots on the SFA, is the low computational cost as well as its clear physical interpretation.

The paper is organized as follows: In Sec. \ref{sec:level2}, we briefly resume the SFA theory and analyze the properties of the temporal integral of the transition matrix.
In Sec. \ref{sec:level2}A we consider LAPE in the streaking regime, i.e., the high frequency pulse is shorter than the IR optical cycle.
In Sec. \ref{sec:level2}B we consider the sideband regime, i.e., the XUV pulse is longer than one IR optical cycle. Finally, in Sec. \ref{sec:level2}C, a train of attosecond pulses is studied.
Concluding remarks are presented in Sec. \ref{conclusions}.
Atomic units are used throughout the paper, except when otherwise stated.

%---------------------------------------------------
\section{\label{sec:level2}Theory and results} 
%---------------------------------------------------

We consider the ionization of an atomic system by the combination of an XUV finite laser pulse assisted by an IR laser, both linearly polarized.
In the single-active-electron (SAE) approximation the time-dependent Schr\"odinger equation (TDSE) reads
\begin{equation}
i\frac{\partial }{\partial t}\left\vert \psi (t)\right\rangle =
\Big[ H_0  + H_\textrm{int}(t)  \Big]
\left\vert
\psi (t)\right\rangle , 
\label{TDSE}
\end{equation}%
where $H_0=\vbp^2/2+V(r)$ is the time-independent atomic Hamiltonian, whose first term
corresponds to the electron kinetic energy, and its second term to the electron-core Coulomb interaction.
The second term in the right-hand side of \eref{TDSE}, i.e.,  
$H_\textrm{int}=\vbr \cdot \vbf_{X}(t) + \vbr \cdot \vbf_{L}(t)$,
describes the interaction of the atom with both time-dependent XUV [$\vbf_{X}(t)$]
and IR [$\vbf_{L}(t)$] electric fields in the length gauge. 

The electron initially bound in an atomic state $|\phi_{i}\rangle$ is emitted to a final continuum state $|\phi_{f}\rangle$, with final momentum $\vbk$ and energy $E=k^2/2$. 
Then, the energy and angle-resolved photoelectron spectra (PES) can be calculated as
\begin{equation}
\frac{\rmd P}{\rmd E \rmd \Omega}=\sqrt{2 E}\ |T_{\infi}|^2 ,
\label{prob1}
\end{equation}
where $T_{\infi}$ is the $T$-matrix element corresponding to the transition $\phi_{i}\rightarrow\phi_{f}$ and $\rmd\Omega=\sin\theta\rmd\theta\rmd\phi$, with $\theta$ and $\phi$ the polar and azimuthal angles of the laser-ionized electron, respectively.

Within the time-dependent distorted wave theory, the transition amplitude in the prior form and length gauge is expressed as
\begin{equation}
T_{\infi}= -i\int_{-\infty}^{+\infty}\rmd t \,\langle\chi_{f}^{-}(\vbr,t)|H_\textrm{int}(\vbr,t)|\phi_{i}(\vbr,t)\rangle,
\label{Tif}
\end{equation}
where $\phi_{i}(\vbr,t)=\varphi_{i}(\vbr)\,e^{i I_{p} t}$ is the initial atomic state, with
ionization potential $I_{p}$, and $\chi_{f}^{-}(\vbr,t)$ is the distorted final state.
\Eref{Tif} is exact as far as the final channel, $\chi_{f}^{-}(\vbr,t)$, is the exact solution of \eref{TDSE},
within the dipole approximation. However, several degrees of approximation have been considered so far to solve
\eref{Tif}. The widest known one is the SFA, which neglects the Coulomb distortion in the final channel
produced on the ejected-electron state due to its interaction with the residual ion and discard the influence of the laser field in the initial ground state~\cite{maciej1994,symphony}. The SFA, for instance, is able to model the 'ring' structures of the above-threshold ionization (ATI) photoelectron spectrum~\cite{maciej1995}.
Hence, we can approximate the distorted final state with a Volkov function, which is the solution of
the TDSE for a free electron in an electromagnetic field
\cite{Volkov}, i.e., $\chi_{f}^{-}(\vbr,t)= \chi_{f}^{V}(\vbr,t)$, where
\begin{eqnarray}
 \chi_{f}^{V}(\vbr,t) &=& 
 (2\pi)^{-3/2} \exp \{ i[\vbk+\vba(t)] \cdot \vbr \}\nonumber \\ 
&&\times \exp{ \frac{i}{2} \int_{t}^{\infty} [\vbk+\vba(t^\prime)]^2 \rmd t^\prime }
\end{eqnarray}
and the vector potential due to the total external field is defined as 
$\vba(t) = -\int_0^t \rmd t'[\vbf_{X}(t')+\vbf_{L}(t')]$.
As the frequency of the XUV pulse is much higher than the IR field one,
and considering the strength of the XUV field is much smaller than the IR one, the XUV contribution to the vector potential can be neglected~\cite{Nagele11, DellaPicca13}. 

With the appropriate choice of the IR and XUV laser parameters, we can assume that the energy domain of the LAPE processes is well separated from the domain of ionization by the IR laser alone.  
In other words, the
contribution of IR ionization is negligible in the energy domain where the absorption of one XUV photon takes place. 
Besides, we set the general expression for the linearly polarized XUV pulse of duration $\tau_X$
as 
\begin{equation}\vbf_{X}(t)=-\hat{\varepsilon}_X F_{X0}(t)\cos(\omega _{X}t),
\end{equation}
where $\hat{\varepsilon}_X$ and $\omega _{X}$ are the respective polarization vector and the carrier frequency of the XUV field. Furthermore, $F_{X0}(t)$ is a nonzero envelope function during the temporal interval $(t_{0},t_{0}+\tau_{X})$ and zero otherwise, that we approximate as its maximum amplitude, i.e.~$F_{X0}(t)\approx F_{X0}$. Thus, the matrix element of \eref{Tif} can be written as
\begin{eqnarray}
T_{\infi} & = & -  \frac{i}{2}  \int_{t_0}^{t_0+\tau_X} F_{X0} \hat{\varepsilon}_X \cdot  \vbd \big[ \vbk+\vba(t)\big] \, \rme^{iS(t)} \,\,\rmd t, \label{Tifg}
\end{eqnarray}
where $S(t)$ is the generalized action 
\begin{equation}
S(t)=-\int_{t}^{\infty}\rmd t'\left[\frac{\big( \vbk+\vba(t')\big)^2}{2} + I_{p} -\omega_{X}\right] ,
\label{S1}
\end{equation}
with the dipole moment defined as $\vbd(\vbv)=  (2\pi)^{-3/2}\langle e^{i \vbv \cdot \vbr} \vert \vbr \vert \varphi_{i}(\vbr) \rangle$. In \eref{Tifg} we have used the rotating wave approximation (RWA) which accounts, in this case, for the absorption of only one XUV photon and neglects, thus,  
the contribution of XUV photoemission.
In addition, during the temporal lapse the XUV pulse is acting,
the IR electric field can be modeled as a cosine-like wave, hence, the
vector potential can be written as
\begin{equation}
 \vba(t)=\frac{F_{L0}}{\omega_L}~\sin{(\omega_{L} t)}~\hat{\varepsilon}_L \label{Avector},
\end{equation}
where $F_{L0}$, $\omega_L$ and $\hat{\varepsilon}_L$ are the peak amplitude, carrier frequency, and polarization vector, respectively.
Considering the $T$-periodicity of the vector potential in \eref{Avector}, i.e., $T=2\pi/\omega_L$, the dipole moment results so, i.e., 
%\begin{equation}
%\vbd \big[ \vbk+\vba(t+nT)\big] =\vbd \big[ \vbk+\vba(t)\big],
%\label{dipN}
%\end{equation} 
%with $n$ a positive integer number, i.e., $n=1,...,N$, and $N=\tau_L/T$. is the total number of cycles comprised in the IR %field. 
%
%\textcolor{red}{(R) en el texto que sigue $N$ esta relacionado más con la duracion del XUV, (por eso aparece en los limites de integracion de la integral de $t_{if}$). Por otro lado, definido como  $N=\tau_L/T$, no estaria bien definido $\vba(t+NT)$ pues seria evaluar $\vba$ fuera del pulso IR. Me parece mejor no entrar en estos detalles y dejarlo como estaba antes:}

\begin{equation}
\vbd \big[ \vbk+\vba(t+NT)\big] =\vbd \big[ \vbk+\vba(t)\big],
\label{dipN}
\end{equation} 
with $N$ a positive integer number.

Let us now analyze some features of the $T$-matrix, \eref{Tif}. To this end we notice that the action $S(t)$ defined in \eref{S1}, %which 
can be written as:
\begin{equation}
S(t)= S_0 + a t + b \cos (\omega_L t) + c \sin (2\omega_L t) \label{S},
\end{equation}
where $S_0$ is a constant that results in a phase that can be omitted and
\begin{eqnarray}
a &=&   \frac{k^{2}}{2}+I_{p}+U_p - \omega_{X}, \nonumber \\
b &=&  -\frac{F_{L0}}{\omega_{L}^{2}}\hat{\varepsilon}_L\cdot \vbk , \label{abc} \\
c &=&-\frac{U_p }{2\omega_{L}}, \nonumber
\end{eqnarray}
where $U_p=\frac{F_{L0}^{2}}{4 \omega_L^2}$ defines the ponderomotive energy.
%
%We then observe that $\tilde{S}(t)=S(t)-a t$ is a time-oscillating function with the same period $T$ of the IR laser, i.e.~
%\begin{equation}
%\tilde{S}(t+n T) = \tilde{S}(t).
%\label{S_N}
%\end{equation}
%\textcolor{red}{(R: esta variable $\tilde{S}$ No se usa en lo que sigue. Creo que no hace falta introducir una nueva variable %solo para estos dos renglones. Lo podriamos dejar como antes:}

\begin{figure}[tbp]
\centering
\includegraphics[angle = 0, width=.45\textwidth]{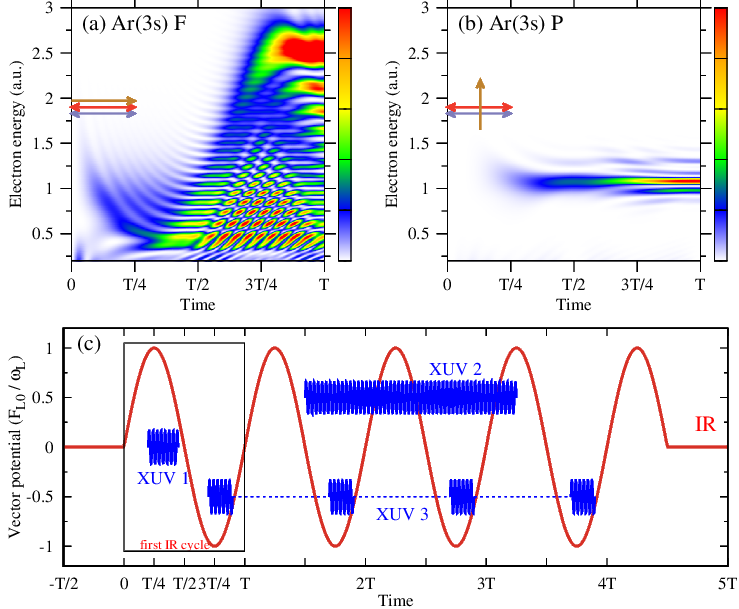}
\caption{
(a)  Squared modulus of the integral $I(t)$, \eref{It}, in arbitrary units, as a function of time and electron energy, for the case of photoionization of Ar($3s$) in forward configuration, i.e.,~the electronic emission direction (yellow arrow) is parallel to both polarization vectors (red -IR- and blue -XUV- horizontal arrows). (b) Idem (a) but for the perpendicular configuration, i.e.,~ the electronic emission is perpendicular to both polarization vectors. 
The IR laser parameters are $F_{L0} = \SI{0.041}{\au}$ and $\omega_L=\SI{0.057}{\au}$, meanwhile for the XUV we take $F_{X0}=\SI{0.01}{\au}$ and $\omega_X=41 \omega_L$. For the IR, these values correspond to a laser intensity and wavelength of $I_L=6\times10^{13}$ W/cm$^2$ and $\lambda_L=800$ nm, meanwhile for the XUV, an intensity and wavelength of $I_X=3.5\times10^{12}$ W/cm$^2$ and $\lambda_X=19.5$ nm, respectively.
(c) Scheme of different XUV+IR photoionization cases (see the text for more details).
}
\label{fig0}
\end{figure}

We then observe that $[S(t)-a t]$ is a time-oscillating function with the same period $T$ of the IR laser field, i.e.~
\begin{equation}
S(t+N T) = S(t) + a N T.
\label{S_N}
\end{equation}
%\textcolor{red}{(R: me parece que escrito asi, despues se entiende mejor las ecuaciones 17, por ejemplo)}
In light of these periodicity properties, Eqs.~(\ref{dipN}) and (\ref{S_N}), we can rewrite the transition matrix, \eref{Tifg}, in terms of the contribution of the first IR cycle only. For that, 
let us introduce the quantity $I(t)$, as the contribution to the transition amplitude from zero to time $t$, i.e.
\begin{equation}
I(t) = \int_{0}^{t}
\ell(t^\prime) \, \rme^{iS(t^\prime)} \, \rmd t^\prime,  \label{It}
\end{equation}
with 
\begin{equation}
\ell(t) =- \frac{i}{2} F_{X0}\hat{\varepsilon}_X \cdot\vbd \big[ \vbk+\vba(t)\big],
\label{lt}  
\end{equation}
providing that $0\leq t\leq T$.
From its proper definition, %in Eq.~(\ref{It}), 
it is clear that $I(t)$ increases from zero at $t=0$ and depends on both the electron energy and the geometrical arrangement between $ \hat{\varepsilon}_X $,  $ \hat{\varepsilon}_L $ and the electron emission direction $ \hat{k}$. 
% We consider that both XUV and IR pulses are linearly polarized in the same direction, i.e., $\hat{\varepsilon}_X = \hat{\varepsilon}_L$.
As an example, in Fig.~\ref{fig0}(a) we show $|I(t)|^2$ for the photoionization of Ar($3s$) in forward ($\hat{k}$ is parallel to both $\hat{\varepsilon}_X$ and $\hat{\varepsilon}_L$) and perpendicular ($ \hat{k}$ is perpendicular to $\hat{\varepsilon}_X$ as well as to $\hat{\varepsilon}_L$) emission configurations (see the arrows in Figs.~\ref{fig0}(a) and~\ref{fig0}(b)). 
Here we consider that both the XUV and IR pulses are linearly polarized in the same direction, i.e., $\hat{\varepsilon}_X = \hat{\varepsilon}_L$.
Figure~\ref{fig0}(c) depicts the different schemes of the LAPE processes studied in this paper.

%---- figura------------------------------------------------------------------------
%--------------------------------------------------------------------------

By performing the transformation $t^\prime=t^{\prime \prime} + NT$, the temporal integral $I(t)$ becomes delayed in $N$ cycles.
Keeping in mind the $T$-periodicity of both $\ell$
and $S$ [see Eqs.~(\ref{S_N}) and~(\ref{lt})], it is straightforward to see that
\begin{eqnarray}
I_1(t)&=&\int_{NT}^{NT+t} \ell(t^\prime) \, \rme^{iS(t^\prime)} \, \rmd t^\prime
\nonumber \\
&=& \int_{0}^{t} \ell(t^{\prime \prime}+NT) \, \rme^{iS(t^{\prime \prime}+NT)}  \rmd t^{\prime \prime} \nonumber \\
&=& I(t ) \, \rme^{iaNT} \label{INT}
\end{eqnarray}
for $t \le T$. We note that when the integrals in Eqs.~(\ref{It}) and~(\ref{INT}) cover a whole IR cycle, they coincide with the laser-assisted photoionization transition matrix for an XUV pulse 
with a duration of one IR cycle
%assisted by an IR single cycle pulse 
[see~\eref{Tifg}]. 
For this reason  we  call $|I(T)|^2 $ as the \emph{intracycle} contribution.

Furthermore, when the XUV pulse covers several IR cycles,
the integral over each cycle can be summed up using \eref{INT} as 
\begin{eqnarray}
I_2 &=& \int_{0}^{NT} \ell(t) e^{iS(t)} \rmd t \nonumber \\\
&=& \sum_{n=0}^{N-1} \int_{nT}^{(n+1)T} \ell(t) e^{iS(t)} \rmd t \nonumber \\
      & = & \sum_{n=0}^{N-1} I(T) e^{ i a n T }\label{vsduration} \\
       &=& I(T) \, \frac{\sin{(a T N /2)}}{\sin{(a T/2)}}\, e^{(i a  T (N-1)/2)}. \nonumber 
\end{eqnarray}
Thus, the PES can be expressed as a product of the \emph{intracycle} factor $|I(T)|^2 $
and the factor $|\sin{(a T N /2)}/\sin{(a T/2)}|^2$, that accounts for the \emph{intercycle} contributions, since it is the result of the phase interference arising from the $N$ different cycles \cite{Arbo2010a,Arbo2010b,Arbo2012}.

The factorization of the transition amplitude in \eref{vsduration} was previously obtained in LAPE~\cite{Gramajo18} and ATI~\cite{Arbo2010a,Arbo2010b,Arbo2012}, within the SCM. In these works, each contribution was recognized as the interference stemming from electron trajectories within the same optical cycle
(intracycle interference) and from trajectories released at different cycles (intercycle interference). 
%In fact, when  $I(T)$ is computed by the saddle point approximation we derive in Eq. (14) of \cite{Gramajo18}.
However, here we prove its validity beyond the SCM as a mere consequence of the periodicity of the transition matrix [see \eref{dipN}]~\cite{proceedingIcpeac19}.

The zeros of the denominator in the intercycle factor, i.e., the energy values satisfying $a T/2 =n\pi$, are avoidable singularities since the numerator also cancels out and maxima are reached at these points. Such maxima are recognized as the \emph{sideband peaks} in the PES. They occur when
\begin{equation}
E_n = n \omega_L + \omega_X -I_p-U_p, \label{sideband}
\end{equation}
corresponding to the absorption (positive $n$) or emission (negative $n$) of $n$ IR photons, following the absorption of one XUV photon.  In fact, when $N\rightarrow \infty$ in \eref{INT}, the intercycle factor becomes a series of delta functions, i.e., $\sum_{n}\delta (E-E_{n})$, satisfying the conservation of energy. 
Instead, for finite XUV pulse duration $\tau_X$ (of the order of $NT$), each sideband peak has a width $\Delta E \sim 2\pi/NT$, fulfilling then the uncertainty relation $\Delta E \tau_X \sim 2\pi$.

Now, we are interested in considering a general situation with arbitrary delays ($t_0$) and XUV durations ($\tau_X$). In order to do so, we express the transition matrix  of \eref{Tifg} in terms of the integral $I(t)$ [\eref{It}].
Therefore, in the following sections we analyze the three different XUV+IR photoionization scenarios sketched in Fig.~\ref{fig0}(c).

%\item[XUV 1]  ----------------------------------------------
\subsection{XUV 1: streaking regime} 
\label{sec:streaking}
%------------------------------------

\begin{figure}[tbp]
\centering
\includegraphics[angle = 0, width=0.5 \textwidth]{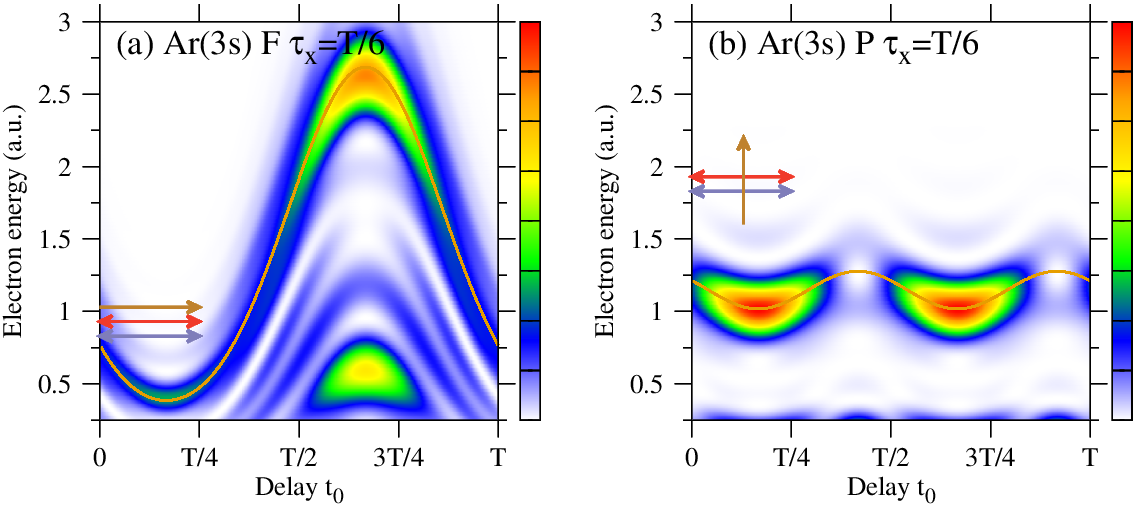}
\caption{PES for an XUV with $\tau_X=T/6$ as a function of the delay $t_0$ for forward (a) and perpendicular (b) configurations.
The orange line corresponds to \eref{KEpar}
%$E(t_0)= [v_0-A_L(t_0+\tau_X/2)]^2/2$ 
in (a) and 
\eref{KEper}
%$E(t_0)= [v_0^2-A_L^2(t_0+\tau_X/2)]/2$ 
in (b).
The laser parameters are the same as those used in Fig.~\ref{fig0}. 
}
\label{fig_streak}
\end{figure}

In the case where the high frequency pulse is shorter than the IR optical cycle [see the XUV 1 scheme in Fig. \ref{fig0} (c)] i.e., $\tau_X < T$,  the integration of the transition matrix from the beginning of the XUV pulse, $t_0$, to its end, $t_0+\tau_X$, in \eref{Tifg} can be written as the subtraction of two integrals in the intervals $[0, t_0+\tau_X]$ and $[0,t_0]$, i.e.
%, the transition matrix can be written as:
%
\begin{eqnarray}
T_{\infi}&=& \int_0^{t_0+\tau_X} \ell(t) e^{iS(t)} \, \rmd t -\int_0^{t_0} \ell(t)  e^{iS(t)} \, \rmd t \label{Eq_streak} \\
      &=& \left\{ \begin{array}{ll}
           I({ t_0+\tau_X})-I({ t_0})                       & \textrm{if $t_0+\tau_X\leq T$} \\
           I({ T}) + I({ t_0 + \tau_X -T})\rme^{iaT} - I({ t_0})  & \textrm{if $ T \leq t_0+\tau_X $} .
                   \end{array} \right.\nonumber
\end{eqnarray}
For simplicity, we have considered the case $t_0\leq T$\footnote{If it is not the case, i.e.~when $t_0=MT+\delta$, we have to insert the factor $\rme^{iaMT}$ before the bracket in \eref{Eq_streak} and replace $t_0$ by $\delta$.}.
Then, taking into account that $I(t)$ is given by \eref{It}, the PES is obtained by inserting \eref{Eq_streak}, into \eref{prob1}, which depends on the delay time $t_0$.
As an illustrative example we show in Fig.~\ref{fig_streak} the %streaking 
PES for Ar($3s$) generated by a short XUV pulse with $\tau_X=T/6$ as a function of  $t_0$, for both the forward [Fig.~\ref{fig_streak}(a)] and perpendicular [Fig.~\ref{fig_streak}(b)] configurations.
 We can observe that the PES for the two cases present the typical streaking pattern~\cite{Mairesse2005,Goulielmakis2004}.
 A simple classical viewpoint considers that the ionization is produced at only one particular instant, corresponding to the stationary time derived from the saddle point equation $dS(t)/dt=0$.
 In this sense, the kinetic energy at that instant of time, which we can adjudicate to the middle of the time interval that the XUV pulse takes action, $t_0 + \tau_X/2$, is
\begin{equation}
  E(t_0)= \frac{\big[ v_0-A_L(t_0+\frac{\tau_X}{2} ) \big]^2}{2}  ,
\label{KEpar}
\end{equation}
where $v_0 = \sqrt{2(\omega_X- I_p )}$ represents the initial classical velocity of the ejected electron. We plot \eref{KEpar} as an orange line. As expected, we observe that the PES for the forward emission configuration follows the shape of the vector potential [\eref{KEpar}] shown in orange line in Fig.~\ref{fig_streak}(a).

The classical viewpoint predicts that the energy maxima in the perpendicular emission occur at 
\begin{equation}
E(t_0)= \frac{\big[ v_0^2-A_L^2(t_0+\frac{\tau_X}{2}) \big]}{2} ,
\label{KEper}
\end{equation}
at the mean time of the XUV pulse~\cite{Goulielmakis2004,Drescher2005}, which is plotted as an orange line in Fig.~\ref{fig_streak}(b).
The PES then oscillates around the classical prediction [\eref{KEper}], as the orange line in Fig.~\ref{fig_streak}(b) illustrates.
We note, however, that there exist some structures in the PES beyond the classical prediction, see e.g.~in Fig.~\ref{fig_streak}(a) at $t_0 \approx 3T/4$ and energy  \SI{0.5}{\au}, that do not strictly represent a classical streaking situation. These structures correspond to the quantum nature of the photoionization phenomenon and stem from the Fourier transform of the XUV squared pulse shape. When we use gaussian or sin$^2$ envelopes instead, these structures vanished (not shown).

%---- figura------------------------------------------------------------------------
%--------------------------------------------------------------------------

\begin{figure}[tbp]
\centering
\includegraphics[angle = 0, width=0.5 \textwidth]{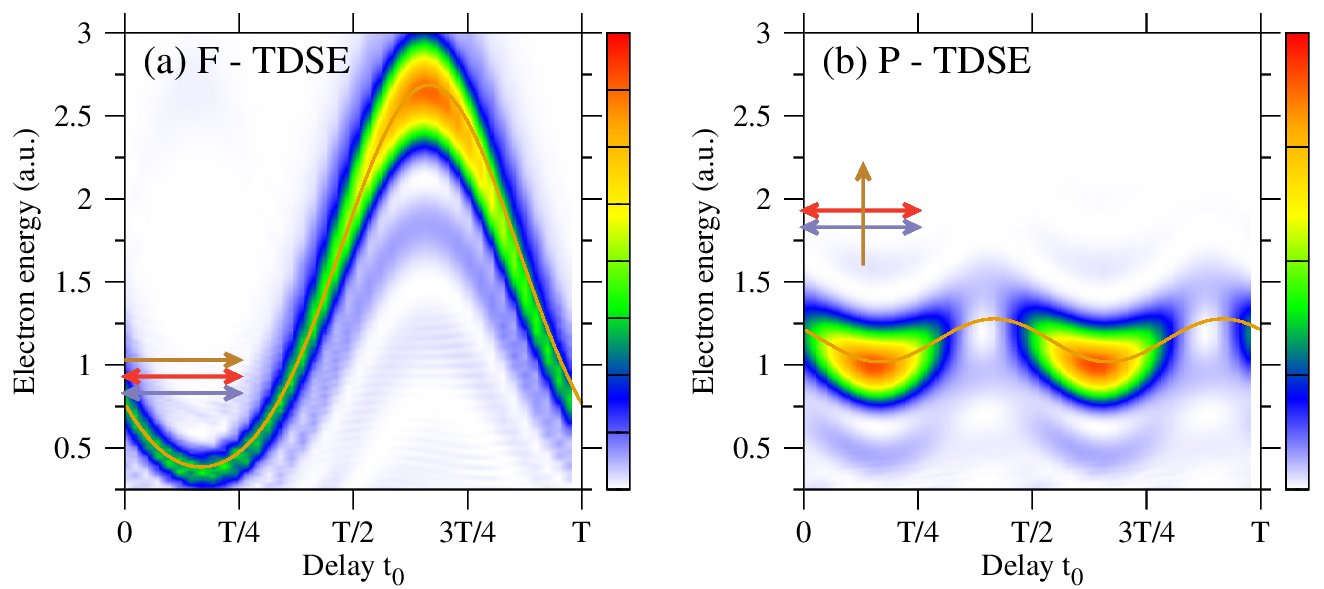}
\caption{Idem as Fig~\ref{fig_streak} but for TDSE results (see the text for more details).}
\label{fig_streak_TDSE}
\end{figure}

In order to corroborate the precedent predictions, we have additionally performed calculations 
by solving \textit{ab initio} the TDSE.
In Fig.~\ref{fig_streak_TDSE} we show the TDSE results for the same field configurations as in Fig.~\ref{fig_streak}. We observe an excellent agreement between both approaches. 
For the  numerical solution of the TDSE we have employed the generalized pseudospectral method combined with the split-operator representation of the time-evolution operator, which was explained
in our previous works \cite{Gramajo16,Gramajo17,Gramajo18}. 
For the computational feasibility of the TDSE calculations, both the XUV and IR fields envelopes are modeled with a trapezoidal shape, comprising one-cycle ramp on and one-cycle ramp off.
%we take the XUV pulse and the IR laser envelope 

%---- figura------------------------------------------------------------------------
%--------------------------------------------------------------------------

%\item[XUV 2] ------------------------------------------ 
\subsection{XUV 2: sidebands regime}
\label{sec:sideband}
%---------------------------------------------------------

When the XUV pulse is longer than one IR period [see the XUV 2 scheme in Fig.~\ref{fig0}(c)] we can sum up the contribution from different cycles as we have presented in \eref{vsduration}. However, in this work,  we are interested in considering 
an arbitrary XUV pulse of duration $\tau_X=NT + \Delta$, that starts at time $t_0 = M T + \delta$, where $\Delta, \delta \leq T$ and $N$, $M$ are integer numbers. Then, using the result introduced in the previous subsection we can write 
\begin{eqnarray}
T_{\infi} &=& \int_{MT + \delta }^{MT+\delta + NT+\Delta} \ell(t) \rme^{iS(t)} \rmd t \nonumber \\
&=&e^{i a T M}\int_{\delta}^{NT + \delta + \Delta}   \ell(t) \rme^{iS(t)} dt \nonumber\\
       &=& e^{i a T M} \left[ \int_0^{NT} \cdots+ \int_{NT}^{NT+\delta+\Delta}\cdots - \int_0^\delta\cdots \right]  \label{general}\\
       &=& e^{i a T M} \left[
 I( T) \frac{\sin{( a T N /2})}{\sin{( a T/2})}  e^{i a  T { (N-1)/2 }} +  \right. \nonumber\\
       && \left. \quad \quad +~ e^{i a  T N} I( \delta+\Delta)- I( \delta)\right] \nonumber
\end{eqnarray}
if $ \delta + \Delta \leq T $, or
\begin{eqnarray}
T_{\infi}&=& e^{i a T M} \left[ I(T) \frac{\sin{( a T (N+1) /2})}{\sin{( a T/2})} e^{i a  T N/2 } + \right.\nonumber\\
       && \left.  + e^{i a  T (N+1)} I( \delta+\Delta-T) - I( \delta)\right] 
       \label{generalp}, 
\end{eqnarray}
if $\delta + \Delta \geq T $.
The transition matrices in Eqs.~(\ref{general}) and (\ref{generalp})
generalize the ones presented in our previous works
\cite{Gramajo16,Gramajo17,Gramajo18,proceedingIcpeac19,Hummert20},
which consider the particular case when the XUV covers an integer number of IR cycles ($\Delta =0$), starting with no delay, i.e., $\delta=0$. In such a case the PES results proportional to
\begin{equation}
|T_{\infi}|^2 =  \underbrace{|I(T)|^{2}}_{\textrm{intracycle}} ~
\underbrace{\left[\frac{\sin{( a T N /2)}}{ \sin{(a T/2) }}\right]^2}_{\textrm{intercycle}}  .
\label{intrainter}
\end{equation} 
This last expression is equivalent to that discussed below \eref{vsduration} and was 
exhaustively studied in Refs.\cite{Gramajo16,Gramajo17,Gramajo18,proceedingIcpeac19,Hummert20}. 
Even though in the general case $\delta$ and $\Delta$ are nonzero, 
we note that the first term inside the brackets in \eref{general}  determines the leading contribution to the 
PES when $N>>1$. This is so due to the increase of the intercycle interference term at $aT=2n\pi$ 
[see the discussion after \eref{sideband}]. In such a case, the PES approximately behaves like \eref{intrainter}.

%---- figura------------------------------------------------------------------------
%\begin{figure}[tbp]
\begin{figure*}
\centering
\includegraphics[angle = 0, width=0.7\textwidth]{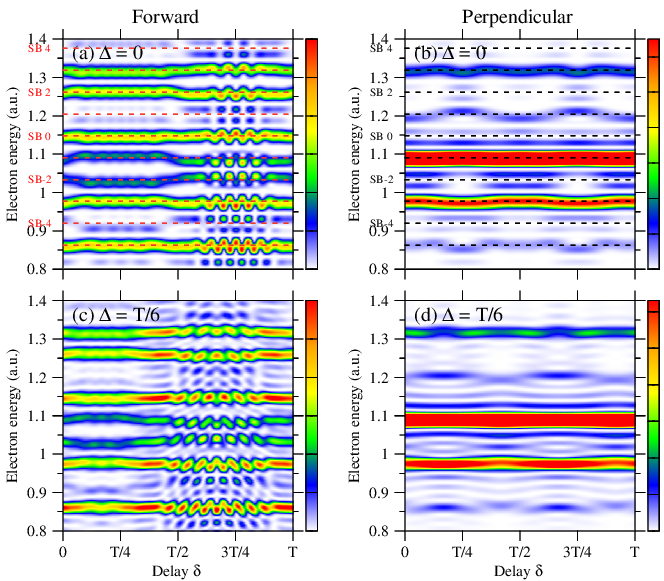}
\caption{
Ar(3s) PES in the sideband regime as a function of $\delta$ (see the text) for forward [(a) and (c)] and  perpendicular [(b) and (d)] emission configurations.   
The XUV pulse has a duration $\tau_X=NT + \Delta$ with $N=2$. In (a) and (b) $\Delta=0$ and in (c) and (d) $\Delta=T/6$.  
The laser parameters are the same as those used in Fig.~\ref{fig0}.
}
\label{fig_sideband}
\end{figure*}
%--------------------------------------------------------------------------

\begin{figure}[tbp]
\centering
\includegraphics[angle = 0, width=0.45 \textwidth]{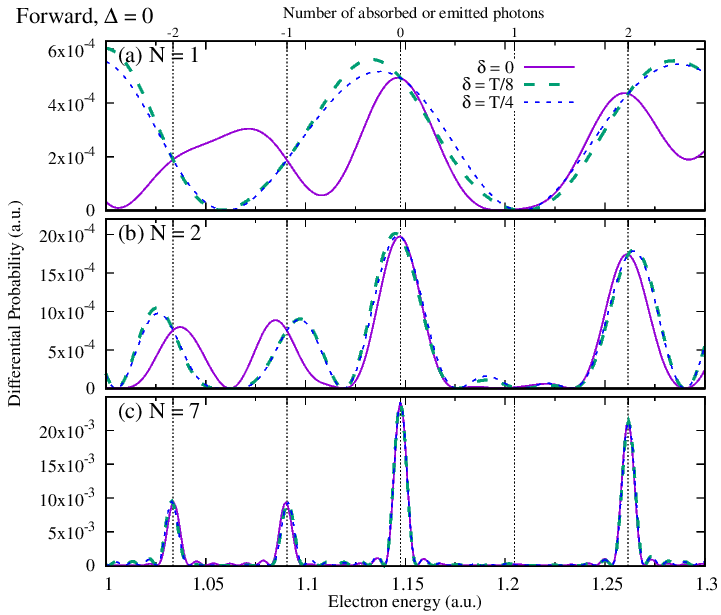} %{Sideband_vs_delta_k}
\caption{PES for Ar(3s), as a function of the electron energy for forward emission and $\Delta = 0$, for three different values of $\delta$ [see \eref{eq_sideband}]. The XUV pulse comprises $N=1$ (a), $N=2$ (b) and $N=7$ (c) IR optical cycles, respectively. 
The laser parameters are kept as those used in the previous figures.  
}
\label{fig_sideband_2}
\end{figure}

In order to study the effect of nonzero $\delta$ and $\Delta$ for finite $N$, we present in Fig.~\ref{fig_sideband} the PES for Ar(3s) using the $T_{\infi}$ of \eref{general}, as a function of the parameter $\delta$,
for both $\Delta = 0$ and $\Delta =T/6$ (in both cases we consider the $I(t)$ of Fig.~\ref{fig0}).
We observe that at the sideband positions (dashed lines) the intensity of the PES remains constant as a function of $\delta$, when $\Delta=0$ [Figs.~\ref{fig_sideband}(a) and~\ref{fig_sideband}(b)] . 
This is so because for $\Delta=0$, according to \eref{general}, we obtain
\begin{eqnarray}
|T_{\infi}|^2 &=& \Big|\frac{\sin{(a T N /2)}}{\sin{(a T/2)}}\Big|^2 \nonumber \\
&\times& \Big|I(T) + I(\delta)\, 2 i\sin(aT/2) \, e^{i a  T /2 } \Big|^2.
\label{eq_sideband}
\end{eqnarray}
Here, at the sideband positions ($aT/2=n\pi$), the second term vanishes and $|T_{\infi}|^2$ results independent of the delay.
This fact can also be observed in Fig.~\ref{fig_sideband_2}, where we show the PES as a function of the electron energy, for different values of $\delta$: at the sideband positions (dashed vertical lines) all the curves agreed each other. Furthermore, the agreement extends to other energies as $N$ increases, where the PES is basically $\delta$ independent [see Fig.~\ref{fig_sideband_2}(c)]. We note that the three curves in Fig.~\ref{fig_sideband_2}(b) correspondingly correlate to cuts of Fig.~\ref{fig_sideband}(a) at $\delta = 0, T/8 $ and $T/4$, respectively.

 Otherwise, in the perpendicular emission case  [Figs.~\ref{fig_sideband}(b) and~\ref{fig_sideband}(d)], there are very little difference in the PES for the two values of $\Delta$ considered. 
 Additionally, the dependence on $\delta$ at the sideband energies is negligible. Hence, generally speaking, the duration and delay of a non-integer number of cycles do not significantly affect the PES.

%---- figura------------------------------------------------------------------------
%--------------------------------------------------------------------------

%---------------------------------
\subsubsection{Integration over the emission directions}
%-----------------------------------

In view of the precedent analysis, the doubly differential PES can
be considered to be approximately proportional to \eref{intrainter},
when the number of IR cycles $N$ is not small.
In this case, we note that the dependence on the emission direction is present only in the intracycle interference factor. This is so because the intercycle factor does not depend on the emission direction (it only relies upon the energy through the factor $a$). As a consequence, the single differential PES ($\rmd P/\rmd E$) can be easily obtained integrating only the intracycle factor, i.e.
\begin{eqnarray}
\frac{\rmd P}{\rmd E} &=&
\sqrt{2E} \int d\Omega |T_{\infi}|^2 \nonumber\\
&=& 
\underbrace{\left[\frac{\sin{( a T N /2)}}{ \sin{(a T/2) }}\right]^2}_{\textrm{intercycle}}  ~
\sqrt{2E} \int d\Omega |I(T)|^{2} . \label{spes}
\end{eqnarray}
 Let us note that not only the double [\eref{intrainter}], but also the single differential PES [\eref{spes}] can thus be written as the product of the intercycle interference factor and the contribution for $N=1$, taking the role of an `intracycle' factor.
In this sense, the factorization of the PES is still valid even for the angular integrated spectra.

%---- figura------------------------------------------------------------------------
\begin{figure}%[tbp]
\centering
\includegraphics[angle = 0, width=0.45 \textwidth]{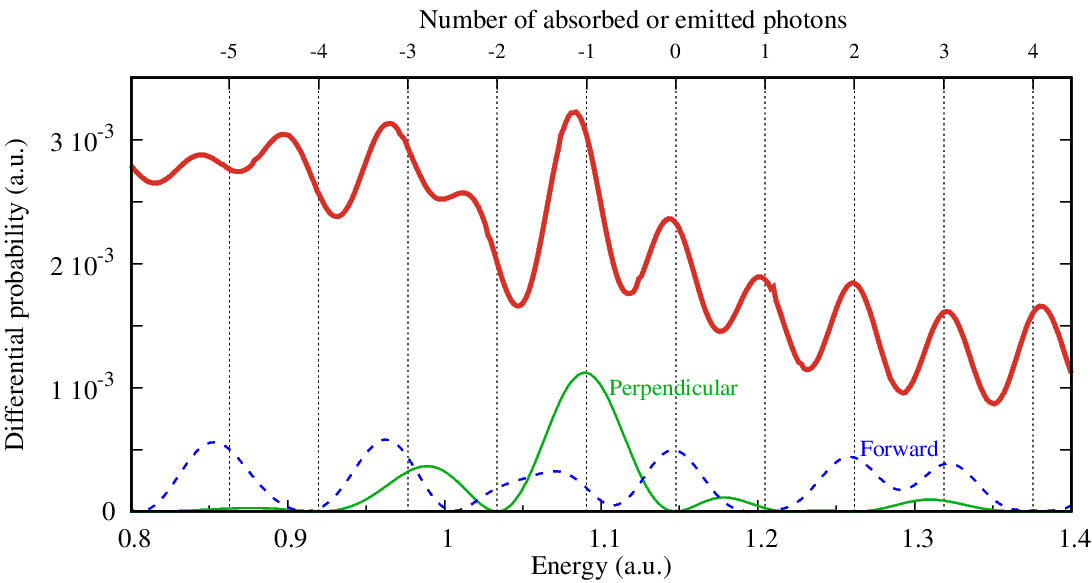}
\caption{PES for Ar(3s), as a function of the electron energy, integrated over all the emission directions. 
Forward and perpendicular emission cases are also shown (see the text for details).
The XUV pulse comprises $N=1$ cycle. The laser parameters are kept as those used in previous figures.  
}
\label{fig_ang_int}
\end{figure}
%--------------------------------------------------------------------------

In  Fig.~\ref{fig_ang_int} we show the single differential PES for Ar(3s) for the case with $N=1$, i.e.,
 $\sqrt{2 E} \int d\Omega |I(T)|^{2}$, as a function of the electron energy. As a reference we also plot the PES for the
 forward and perpendicular emission directions of Figs.~\ref{fig0}(a) and~\ref{fig0}(b), respectively, corresponding to cuts at $t=T$, and multiplied by $\sqrt{2E}$.
 We observe that the angular integrated PES (red thick line) presents several peaks, that do not necessarily match at the sideband positions (vertical dashed lines).
 However, when the XUV is longer than one IR cycle, the intercycle factor must be considered. Then, the PES presents thus maxima at the sideband peaks positions but modulated by this red thick curve (not shown).

%---------------------------------------------
\subsubsection{Intra- and inter half-cycle interferences}
%---------------------------------------------
In this subsection we
consider the particular situation when the electron emission direction is perpendicular to the IR laser polarization vector, $\hat{\varepsilon}_L \perp \vbk$. 
Because of this configuration, $b=0$ in \eref{S} and $[S(t)-at]$ has not only $T$- but also a $T/2$-periodicity. Besides, we also consider that the dipole element also satisfies 
\begin{equation}
\hat{\varepsilon}_X \cdot \vbd[\vbk+\vba(t+T/2)] = \pm \hat{\varepsilon}_X \cdot\vbd[\vbk+\vba(t)],  \label{dipole+-}
\end{equation}
 i.e., it is symmetric or antisymmetric with respect to the middle of the IR cycle. Under these circumstances, the integral $I(t)$ of \eref{It} over one IR cycle can be 
 written as
\begin{eqnarray}
I(T) &=& \int_0^{T/2} \ell(t) e^{iS(t)} dt + \underbrace{\int_{T/2}^T \ell(t) e^{iS(t)} dt}_{\pm e^{iaT/2} I(T/2)} \\
     &=& I(T/2)(1 \pm e^{iaT/2}), \nonumber
  \end{eqnarray}  
where we have split $I(T)$ as a sum over the two IR half cycles.
Then, depending on the symmetric ($+$) or antisymmetric ($-$) character of the dipole element with respect to $T/2$ we have,
\begin{eqnarray}     
|I(T)| &=& |2 ~ I(T/2) ~ \cos{(a T/4)} | \, \, \textrm{ {\scriptsize if} }+  \, \, \textrm{ {\scriptsize (symmetric)} } \label{si} \\
|I(T)| &=& |2 ~ I(T/2) ~ \sin{(a T/4)} | \, \, \textrm{ {\scriptsize if} }- \, \, \textrm{ {\scriptsize(antisymmetric).} } \label{asi}
\end{eqnarray}  
The factor $\cos(a T/4)$ [$\sin(a T/4)$] in \eref{si} [\eref{asi}], cancels out odd (even) sideband peaks in the intercycle contribution. 
As a consequence, the PES presents structures corresponding to the absorption or emission of only an even (symmetric dipole element) or odd (antisymmetric dipole element) number of IR photons. Furthermore, the energy difference between two consecutive sideband peaks is $2\omega$ instead of $\omega$, as the general conservation energy rule in \eref{sideband} indicates.
 
In the present work, we consider the Ar($3s$) dipole element from
a hydrogen-like excited state, i.e.,~\eref{3s}, evaluated at $\vbv = \vbk + \vba(t)$. Since both the XUV and IR laser pulses have the same polarization direction, the dipole element in the perpendicular emission case results antisymmetric, i.e., the $(-)$ instance in \eref{dipole+-} should be used.
For antisymmetric dipole elements the %PES
$|T_{\infi}|^2$
of \eref{intrainter} becomes
\begin{eqnarray}
|T_{\infi}|^2&=&  \underbrace{4 ~\underbrace{|I(T/2)|^{2}}_{\textrm{intrahalfcycle}}~ \sin^2(aT/4) }_{\textrm{intracycle}}~
\underbrace{\left[\frac{\sin{( a T N /2)}}{ 2 \sin{(a T/4) \cos(a T /4)}}\right]^2}_{\textrm{intercycle}}
\label{PEasi-1} \nonumber\\
&=& \underbrace{|I(T/2)|^{2}}_{\textrm{intrahalfcycle}}~\underbrace{\left[\frac{\sin{( a T N /2)}}{ \cos(a T /4) }\right]^2}_{\textrm{interhalfcycle}}, \label{PEasi}
\end{eqnarray}
which  reaches  maxima only for odd $n$ and it becomes suppressed at energy values $E_n$ with even $n$ (see \eref{sideband}).
In particular, the absorption of only one XUV photon alone (in the absence of absorption or
emission of IR photons) is forbidden.

In Figs.~\ref{fig_sideband}(b) and \ref{fig_sideband}(d) we  note 
that the emission probability vanishes along the dashed lines marked as SB0, SB2, SB-2, etc. This absence of even order sideband peaks  confirms indeed the selection rule that determines the presence of only odd sideband orders even for non-zero $\delta$ and $\Delta$ values.
We can also observe that, effectively, in %Fig.~\ref{fig_train}(b)
Fig.~\ref{fig_ang_int} the  intracycle factor $I(T)$ 
for perpendicular emission direction (green solid line) vanishes at even sideband positions according to \eref{asi}. 

Alternatively, for symmetric dipole elements, the odd sideband orders cancel out whereas the even orders stay put  \cite{proceedingIcpeac19}.
In correspondence with our previous analysis within the SCM (see Eq.~(18) in Ref.~\cite{Gramajo17}), 
\eref{PEasi-1} indicates that the PES can be factorized into two different ways:
(i)  as the product of intra- and intercycle interference factors and
(ii) as the product of intrahalf- and interhalf-cycle interference contributions.
Obviously, the two different factorizations give rise
to the same results. 

%-----------------------------------------------------
\subsection{XUV 3: attosecond pulse train}
\label{train}
%---------------------------------------------------------

We study the LAPE process for the case of a train of $J$ identical (in phase) pulses of duration $\tau_X$ each [see the XUV 3 scheme in Fig.~\ref{fig0}(c)]. Each pulse is repeated every $D$ cycles (clearly $\tau_X \le DT$), where $D$ is any positive integer number and the $j$-th pulse starts at $t_{0j}=t_0+(j-1) D T$ with $j=1,...,J$. Then, the temporal integral of the transition matrix in \eref{It} becomes a sum of $J$ integrals over the temporal intervals $[t_{0j}, t_{0j} + \tau_X]$, where the $j$-th XUV pulse acts,
\begin{eqnarray}
T_{\infi}        &=&\sum_{j=1}^J T_{\infi}^{(j)}, \label{sumTifj}
\end{eqnarray}
where each individual transition matrix $T_{\infi}^{(j)}$ corresponds to the $j$-th pulse and is given by
\begin{eqnarray}
    T_{\infi}^{(j)} &=& \int_{t_{0j}}^{t_{j0}+\tau_X} \ell(t) ~ \rme^{iS(t)} \rmd t \nonumber \\
    &=& \rme^{ia(j-1)DT}  T_{\infi}^{(1)}.
\end{eqnarray}
Following the same reasoning as in \eref{vsduration}, and using \eref{sumTifj}, we find that
\begin{eqnarray}
T_{\infi} & = & \sum_{j=1}^J~\rme^{ia(j-1)DT}  T_{\infi}^{(1)} =  T_{\infi}^{(1)}  \sum_{j=0}^{J-1}~\rme^{iaDT j}  \nonumber \\
       & = &  T_{\infi}^{(1)}  ~ \rme^{[i D  (J-1) aT /2]} ~\frac{\sin(J D aT/2) }{\sin( D a T/2)}. \nonumber
\end{eqnarray}
We can thus finally write
\begin{eqnarray}
|T_{\infi}|^2 & = &  \underbrace{|T_{\infi}^{(1)}|^{2}}_{\textrm{intrapulse}} ~
\underbrace{\left[ \frac{\sin(J D a T/2) }{\sin( D a T/2)}\right]^2}_{\textrm{interpulse}}.
\label{intrainterPULSE}
\end{eqnarray}
As in the previous cases, we can split up $|T_{\infi}|^2$ as a product of two interference factors: the \textit{intrapulse} interference, that corresponds to the emission probability of an isolated pulse, and the \textit{interpulse} interference, that accounts for the interference due to the coherent emission from different pulses.

%---- figura------------------------------------------------------------------------
\begin{figure}[tbp]
\centering
\includegraphics[angle = 0, width=0.45 \textwidth]{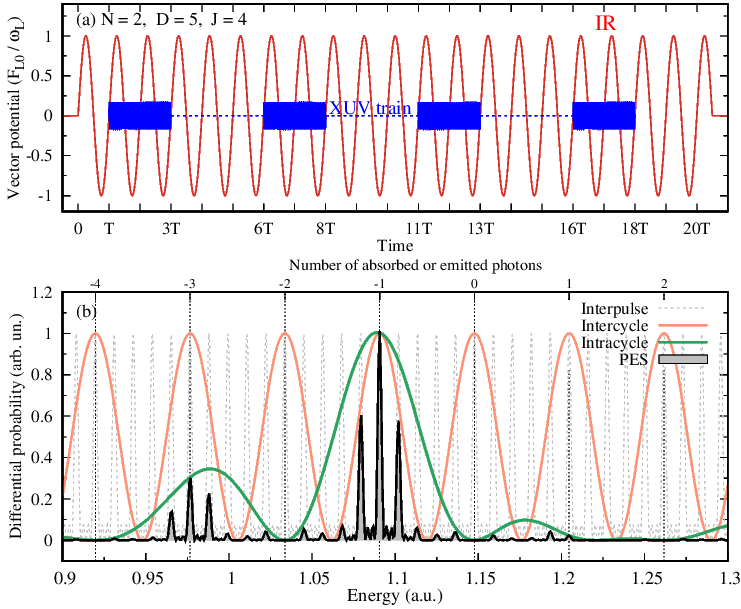}
\caption{(a) Pulse train with four pulses of duration $\tau_X = N T$ with $N=2$ and repetition rate $D T$ with $D=5$.
(b) PES for perpendicular emission configuration. 
The laser parameters are the same as those used in previous figures.
%$F_{L0} = 0.041$ a.u., $\omega_L=0.057$ a.u., $F_{X0}=0.01$ a.u. and $\omega_X=41 \omega_L$. % and $\tau_X=3 T$.  
}
\label{fig_train}
\end{figure}
%--------------------------------------------------------------------------

In \eref{intrainterPULSE}, the intrapulse factor can be calculated considering the theory explained in Secs. \ref{sec:streaking} or \ref{sec:sideband}, i.e., $T_{\infi}^{(1)}$ must be replaced by \eref{Eq_streak}, $I(T)$ or \eref{general}, depending on the case.
For instance, if the repetition rate of the pulse train is every cycle ($D=1$, as in the case plotted in Fig.~\ref{fig0}(c)), the interpulse factor looks exactly like the intercycle interference one. Then, $T_{\infi}^{(1)}$ corresponds to \eref{Eq_streak}.
For the special case that $\tau_X = T$, the transition probability \eref{intrainterPULSE} %is 
becomes
equal to \eref{intrainter}, considering that the number of IR optical cycles covered by the XUV pulse,
%of the XUV, 
$N$, is equal to $J$, the number of XUV pulses. 

For $D > 1$ we observe that the interpulse factor reaches maxima every time the denominator is zero, i.e., at $aT/2 = m \pi /D$.
This means that there are $D-1$ `interpulse' secondary peaks
between two consecutive sidebands (defined by \eref{sideband}) or, likewise, when 
\begin{equation}
E_m = m \frac{\omega_L}{D}  + \omega_X - I_p - U_p .
\label{D_peaks}
\end{equation}
The energy values in \eref{D_peaks} can be understood as a particular energy conservation law for the exchange of $m$ photons after the absorption of one single XUV photon, where its energy results a fraction of the IR photon energy, i.e., $\omega_L/D$. We note that for $D=1$ the interpulse and intercycle (or sidebands) peaks agree.

As an example, we show in Fig.~\ref{fig_train}(a) the temporal profile of a pulse train consisting of four XUV identical pulses, each of duration twice the IR optical cycle, i.e., $\tau_X = 2 T$, and a periodicity of five IR cycles. In Fig.~\ref{fig_train}(b) we depict the PES corresponding to the electron emission from Ar(3s) in the perpendicular direction (black curve), which can be regarded as the multiplication of several factors: (i) the intracycle factor $|I(T)|^2$ (green curve), which is a cut of the Fig.~\ref{fig0} at $t=T$, (ii) the intercycle factor of \eref{intrainter} with $N=2$ (orange curve), and (iii) the interpulse factor (grey dashed curve), which presents four narrow peaks between two consecutive sidebands. We must point out that the even sidebands vanish in the intracycle factor as a consequence of destructive intra-half-cycle interference, as discussed before. 

%\textcolor{red}{LO QUE SIGUE QUIZAS PODEMOS SACARLO Y QUE QUEDE PARA OTRO PAPER. }
%\textcolor{red}{Falta explicar lo que quiere decir counterphase. Habr\'ia que decir que el caso anterior es 'in phase'?}

Another interesting case is a pulse train of counterphase pulses, i.e., identical pulses but with a phase that changes in $\pi$ between consecutive pulses. This kind of pulse train is created where only odd harmonics of a given monochromatic field are used to generate it~\cite{JimenezGalan2013}. 
Since the transition matrix, \eref{Tifg}, is proportional to the amplitude of the XUV pulse, we only need to add a factor $(-1)^j$ in each term of the sum, \eref{sumTifj}, to calculate the transition matrix,
\begin{equation}
T_{\infi}  =   T_{\infi}^{(1)}  \sum_{j=0}^{J-1}~\rme^{i(aDT +\pi) j } .
\end{equation}
Therefore, we can write
\begin{equation}
|T_{\infi}|^2  =  |T_{\infi}^{(1)}|^{2}  ~
\left[ \frac{\sin(J D a T/2 + J\pi/2) }{\sin( D a T/2 + \pi/2)}\right]^2.
\label{intrainterPULSE2}
\end{equation}
This equation shows that interpulse peaks show up when $D a T/2 +\pi/2= m\pi$, which means that there are $D$ secondary peaks in between two consecutive sidebands. They appear at energies
\begin{equation}
E_m = \left(m-\frac{1}{2}\right) \frac{\omega_L}{D} + \omega_X -I_p-U_p .
\label{NOTsideband}
\end{equation}
Thus, the position of the peaks appearing from ionization due to a train of counterphase pulses are shifted with respect to the position of the 'in phase' secondary peaks, \eref{D_peaks}, by an energy equal to $\omega_L/2D$.

\begin{figure}[btp]
\centering
\includegraphics[angle = 0, width=0.45 \textwidth]{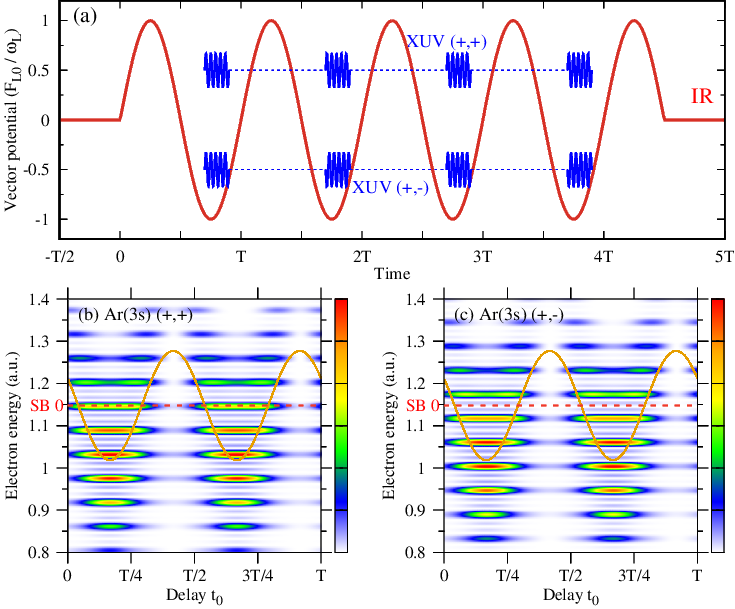}
\caption{(a) XUV pulse train with $D=1$ and four pulses each of duration $\tau_X=T/6$. The $(+,+)$  indicates an 'in phase' pulse train and $(+,-)$ indicates a counterphase one. 
(b) Ar($3s$) PES for perpendicular emission configuration, using the $(+,+)$ pulse train as a function of the delay $t_0$, and 
(c) the same as (b) but for the $(+,-)$ pulse train. 
The red dashed line corresponds to the zero-order sideband and the orange solid line represents the expected streaking energy, \eref{KEper}.
The laser parameters are the same as in Fig.~\ref{fig_train}.
}
\label{fig_train_d1}
\end{figure}

In the following, we consider the particular case of a repetitiveness of one IR cycle, i.e., $D=1$, and compare the ionization probability of Ar(3s) for both the in phase and counterphase pulse train cases.
In the Fig.~\ref{fig_train_d1}(a) we show the time profile of two XUV pulse trains. The upper one, labeled as $(+,+)$, is composed of four identical pulses, whereas the lower one, labeled as $(+,-)$, has alternating zero and $\pi$ phases. This means that the first and third pulses have opposite sign with respect to the second an fourth ones.
The corresponding perpendicular emission PES are shown in Figs.~\ref{fig_train_d1}(b) and~\ref{fig_train_d1}(c), which result from the product of the spectrum shown in Fig.~\ref{fig_streak}(b) and the respective interpulse factor of \eref{intrainterPULSE} or \eref{intrainterPULSE2} for the in phase and counterphase cases, respectively. We include (orange line) in Figs.~\ref{fig_train_d1}(b) and~\ref{fig_train_d1}(c) the expected streaking energy, \eref{KEper}.
The horizontal red dashed line indicates the position of the sideband of order $n=0$ (SB0). 
Note that in Fig.~\ref{fig_train_d1}(c) there is no coincidence between the peaks and the sideband positions because the interpulse peaks in the counterphase case [\eref{NOTsideband}] are shifted with respect to the 'in phase' one [\eref{D_peaks}].

%---- figura------------------------------------------------------------------------
%--------------------------------------------------------------------------

%---------------------------------------------------

\section{Conclusions}
\label{conclusions}
%---------------------------------------------------

We have studied the electron emission produced by an XUV pulse assisted by an IR laser field, emphasizing the analytical properties
inferred from the SFA transition matrix element. 
We have covered a broad range of LAPE situations: both the streaking and sideband regimes for a isolated attosecond pulse, as well as the case of a pulse train.
In all these cases, we have found that the integral of the transition amplitude over the time the XUV pulse acts, can be written as a function of a kernel $I(t)$, defined for only one IR cycle.
With this quantity, the PES can be easily build up
for several configurations of the XUV+IR fields.
Furthermore, we note that our scheme can be applied not only within the SFA but also to other more elaborated approaches, e.g.~the Coulomb-Volkov approximation, as long as the dipole element $\vbd[\vbk+\vba(t)]$ maintains the $T$-periodicity with respect to the IR laser field and the depletion of the ground state is negligible.

In particular, for the case of LAPE due to a pulse train, we have shown that not only intra-, inter-, intrahalf- and interhalfcycle interferences arise, but also intra- and interpulse interference contributions are present as a direct consequence of the periodicity and symmetry of the transition matrix element. All these interference factors manifest themselves as recognizable structures in the PES and would allow to extract structural information from the target system.

%---------------------------------------------------
\appendix
\section{Dipole element}\label{apen_dipol}
%---------------------------------------------------

The dipole transition element is defined as
\begin{equation}
\vbd_i(\vbv) = \frac{1}{(2\pi)^{3/2}} \int \rmd \vbr \, \exp[-i \vbv\cdot \vbr]\,  \vbr \, \phi_i(\vbr),
\end{equation}
where $\phi_i$ is a hydrogen-like bound state. For the case of a hydrogenic $3s$ state we can write
\begin{eqnarray}
\vbd_{3s}(\vbv) &=& -\frac{i}{\pi}\, 2^{7/2}\alpha^{5/2}\, 
\frac{ \vbv}{ (v^2+\alpha^2)^5} \,\Big(3 v^4 +11\alpha^4 - 18 \alpha^2 v^2 \Big), \label{3s}\nonumber\\
\end{eqnarray}
%\\
%\vec{d}_{3p0}(\vec{v}) &=& \frac{\sqrt{3}}{\pi}\, 2^{4} \alpha^{7/2}\,\frac{ 1}{ (v^2+\alpha^2)^5} \,
%\Big[  (v^4-\alpha^4)\, \hat{z} -2 (3 v^2 -5 \alpha^2) (\vec{v}\cdot \hat{z})\, \vec{v}\Big] \label{d3p0} \\
%\vec{d}_{3p1}(\vec{v}) &=& \frac{\sqrt{3}}{\pi}\, 2^{7/2} \alpha^{7/2}\,\frac{ 1}{ (v^2+\alpha^2)^5} \times \nonumber \\
% &&
%\Big[  (3 v^2-5 \alpha^2)
%\left( \begin{array}{c} (v_x+iv_y)^2 - v_z^2 \\ -i(v_x+iv_y)^2 -iv_z^2 \\ 2 v_z(v_x+iv_y) \end{array} \right) + (\alpha^4 +2v^4-5\alpha^2 v^2)\left( \begin{array}{c} 1\\i\\0 \end{array} \right)
%\Big]
%\end{eqnarray}
where $\alpha = \sqrt{2I_p}$.
We have considered the ionization energy %$I_p =15.76$ eV for the $3p$ subshells and 
$I_p =27.623$ eV (1.015 a.u.) for the $3s$ state of Ar. 
Separating the $v^2$ dependence, the $\hat{z}$-component of the dipole element can be reduced to
\begin{eqnarray}
\vbd_{3s} \cdot\hat{z} &=&            v_z          f_1(v^2),
\end{eqnarray}
where we have introduced the function $f_1$, to explicitly indicate the dependence on the modulus squared of its variable $\vbv$, i.e.
%$f_j$ (for $j=1$ to 4) to indicate explicitly the dependence on   the modulus of the vector $\vec{v}$:
\begin{eqnarray}
f_1(v^2) &=&       -i \frac{\, 2^{7/2}\alpha^{5/2}\, (3 v^4 +11\alpha^4 - 18 \alpha^2 v^2 )}{\pi(v^2+\alpha^2)^5}.
\end{eqnarray}

%\nonumber \\
%\vec{d}_{3p0}\cdot\hat{z} &=& f_2(v^2) + v_z^2        f_3(v^2) \nonumber \\
%\vec{d}_{3p1}\cdot\hat{z} &=&            v_z(v_x+iv_y)f_4(v^2) \nonumber
%\end{eqnarray}
%where we have introduced the functions 
%$f_j$ (for $j=1$ to 4) to indicate explicitly the dependence on   the modulus of the vector $\vec{v}$:
%\begin{eqnarray}
%f_1(v^2) &=&       -i \, 2^{7/2}\alpha^{5/2}\, (3 v^4 +11\alpha^4 - 18 \alpha^2 v^2 )
%                                                                  / \pi(v^2+\alpha^2)^5        \nonumber  
%\end{eqnarray}
%\nonumber \\
%f_2(v^2) &=& \sqrt{3} \, 2^{4}  \alpha^{7/2}\, (v^4-\alpha^4 )    / \pi  (v^2+\alpha^2)^5 
%\nonumber \\
%f_3(v^2) &=&-\sqrt{3} \, 2^{5}  \alpha^{7/2}\, (3 v^2 -5 \alpha^2)/ \pi  (v^2+\alpha^2)^5 
% \nonumber \\
%f_4(v^2) &=& \sqrt{3} \, 2^{9/2}\alpha^{7/2}\, (3 v^2-5 \alpha^2) / \pi  (v^2+\alpha^2)^5
% \nonumber
%\end{eqnarray}

\begin{acknowledgments}
This work is supported by CONICET PIP 11220130100386CO, PICT 2016-0296 and PICT-2017-2945 of the ANPCyT
and 06/C033 (2019-2020) UNCuyo
(Argentina). We acknowledge the Spanish Ministry MINECO (National Plan
15 Grant: FISICATEAMO No. FIS2016-79508-P, SEVERO OCHOA No. SEV-2015-0522, FPI), European Social Fund, Fundació Cellex, Fundaci\'o Mir-Puig, Generalitat de Catalunya (AGAUR Grant No. 2017 SGR 1341, CERCA/Program), ERC AdG NOQIA, EU FEDER, European Union Regional Development Fund-ERDF Operational Program of Catalonia 2014-2020 (Operation Code: IU16-011424), MINECO-EU QUANTERA MAQS (funded by The State Research Agency (AEI) PCI2019-111828-2 / 10.13039/501100011033), and the National Science Centre, Poland-Symfonia Grant No. 2016/20/W/ST4/00314. 
\end{acknowledgments}

%---------------------------------------------------
%\bibliographystyle{arthur}
\bibliography{biblio}

%\nocite{*} 
% Produces the bibliography via BibTeX.
%---------------------------------------------------

\end{document}